\documentclass[a4paper,11pt]{article}
\usepackage{jinstpub} 
\usepackage{siunitx}
\DeclareSIUnit{\sample}{Sa}
\DeclareSIUnit{\pixel}{pixel}
\usepackage{subcaption}
\usepackage{multirow}
\usepackage{graphicx}
\usepackage{subcaption}

\title{\boldmath SiPM optical readout of a PICOSEC Micromegas precise timing detector}
\author[a]{F.M. Brunbauer}
\author[b]{R. Aleksan}
\author[b]{S. Aune}
\author[c]{J. Bortfeldt}
\author[d]{A. Breskin}
\author[e,f]{M. Brunoldi}
\author[g]{J. Datta}
\author[h]{G. Fanourakis}
\author[a,i]{K.J. Floethner}
\author[j]{M. Gallinaro}
\author[k]{F. Garcia}
\author[b]{I. Giomataris}
\author[l]{F. Guerra}
\author[m]{D. Janssens}
\author[a,n]{E. Jelinkova}
\author[b]{A. Kallitsopoulou}
\author[o]{I. Karakoulias}
\author[p]{M. Kovacic}
\author[b]{P. Legou}
\author[a]{M. Lisowska}
\author[q]{J. Liu}
\author[i,r]{M. Lupberger}
\author[a]{D.J.G. Marques}
\author[a,r]{H. Muller}
\author[a]{E. Oliveri}
\author[b]{T. Papaevangelou}
\author[s]{M. Pomorski}
\author[a]{L. Ropelewski}
\author[o,t]{D. Sampsonidis}
\author[a]{T. Schneider}
\author[a,u]{B. Schoenfelder}
\author[s]{E. Scorsone}
\author[a]{M. van Stenis}
\author[v]{Y. Tsipolitis}
\author[o,t]{S. Tzamarias}
\author[w]{A. Utrobicic}
\author[e,f]{I. Vai}
\author[a]{R. Veenhof}
\author[e,f]{P. Vitulo}
\author[q]{X. Wang}
\author[a,x]{S. White}
\author[y]{D. Zavazieva}
\author[q]{Z. Zhang}
\author[q]{Y. Zhou}

\affiliation[a]{{European Organization for Nuclear Research (CERN), 1211 Geneve 23, Switzerland}}

\affiliation[b]{{IRFU, CEA, Université Paris-Saclay, F-91191 Gif-sur-Yvette, France}}

\affiliation[c]{{Department for Medical Physics, Ludwig Maximilian University of Munich, Am Coulombwall 1, 85748 Garching, Germany}}

\affiliation[d]{{Weizmann Institute of Science, Herzl St 234, Rehovot, Israel}}

\affiliation[e]{{Dipartimento di Fisica, Università di Pavia, Via Bassi 6, 27100 Pavia, Italy}}

\affiliation[f]{{INFN Sezione di Pavia, Via Bassi 6, 27100 Pavia, Italy}}

\affiliation[g]{{Department of Physics and Astronomy, Stony Brook University, Stony Brook, NY 11794-3800, USA}}

\affiliation[h]{{Institute of Nuclear and Particle Physics, NCSR Demokritos, GR-15341 Agia Paraskevi, Attiki, Greece}}

\affiliation[i]{{Helmholtz-Institut für Strahlen- und Kernphysik, University of Bonn, Nußallee 14–16, 53115 Bonn, Germany}}

\affiliation[j]{{Laboratório de Instrumentacão e Física Experimental de Partículas, Lisbon, Portugal}}

\affiliation[k]{{Helsinki Institute of Physics, University of Helsinki, FI-00014 Helsinki, Finland}}

\affiliation[l]{{Department of Physics, University of Turin, 10125 Torino, Italy}}

\affiliation[m]{Inter-University Institute for High Energies (IIHE), Université libre de Bruxelles, Boulevard du Triomphe, CP 230, 1050 Brussels, Belgium}

\affiliation[n]{{Czech Technical University in Prague, Jugoslávských partyzánů 1580/3, 160 00 Praha 6-Dejvice, Czechia}}

\affiliation[o]{{Department of Physics, Aristotle University of Thessaloniki, University Campus, GR-54124, Thessaloniki, Greece}}

\affiliation[p]{{University of Zagreb, Faculty of Electrical Engineering and Computing, 10000 Zagreb, Croatia}}

\affiliation[q]{{State Key Laboratory of Particle Detection and Electronics, University of Science and Technology of China, Hefei 230026, China}}

\affiliation[r]{{Physikalisches Institut, University of Bonn, Nußallee 12, 53115 Bonn, Germany}}

\affiliation[s]{{CEA-LIST, Diamond Sensors Laboratory, CEA Saclay, F-91191 Gif-sur-Yvette, France}}

\affiliation[t]{{Center for Interdisciplinary Research and Innovation (CIRI-AUTH), Thessaloniki 57001, Greece}}

\affiliation[u]{{Fakultät für Physik, Technische Universitaet Wien, Karlsplatz 13, 1040 Vienna, Austria}}

\affiliation[v]{{National Technical University of Athens, Athens, Greece}}

\affiliation[w]{{Ruđer Bošković Institute, Bijenička cesta 54., 10 000 Zagreb, Croatia}}

\affiliation[x]{{University of Virginia, USA}}

\affiliation[y]{{Technion - Israel Institute of  Technology, Haifa, 3200003, Israel}}

\emailAdd{florian.brunbauer@cern.ch}

\abstract{
By combining a Cherenkov radiator with a semi-transparent photocathode and a two-stage Micromegas amplification element, PICOSEC Micromegas detectors have achieved \SI{<20}{\pico\second} charged-particle time resolution with electronic readout of induced signals. Building upon this performance, we describe SiPM-based optical readout as a novel technique for PICOSEC Micromegas detectors. Single-channel detector prototypes based on glass Micromegas technology were read out with an array of SiPMs. To allow operation in a strongly quenched gas mixture, a solid wavelength shifter is integrated in the glass Micromegas to convert UV scintillation to the visible range and maximise scintillation light output.
Timing performance of glass Micromegas in electronic and optical readout modes is compared and \SI{<150}{\pico\second} time resolution is achieved by optically read out scintillation light signals. The achievable spatial resolution from centre-of-gravity reconstruction of event positions from the amplitudes of signals on multiple SiPM channels is $\sigma = \SI{1.25}{\milli\metre}$. This demonstrates the possibility of dual-readout of PICOSEC glass Micromegas: electronic signals and optical signals. Finally, camera-based optical readout is presented to demonstrate imaging as another avenue for position reconstruction.

}


\begin{document}

\maketitle

\flushbottom

\section{Introduction}
Precise charged particle timing is an important tool for time-of-flight measurements, particle identification and the reconstruction of interactions in particle and nuclear physics experiments. Detectors with time resolutions of a few tens of picoseconds can improve the separation of closely spaced events and extend the capabilities of systems combining tracking and timing measurements.
The PICOSEC Micromegas concept addresses these requirements by combining a Cherenkov radiator with a semi-transparent photocathode and a two-stage gaseous amplification structure~\cite{Bortfeldt2018Picosec}. Cherenkov photons produced by a traversing charged particle release photoelectrons from the photocathode, which undergo pre-amplification in a thin drift region followed by further amplification in the Micromegas amplification gap. Localising the production of the primary electrons at the photocathode reduces the timing fluctuations associated with the distribution of primary ionisation along the particle trajectory in conventional gaseous detectors. The first PICOSEC Micromegas prototype demonstrated a time resolution of \SI{24}{\pico\second} for \SI{150}{\giga\electronvolt} muons~\cite{Bortfeldt2018Picosec}. More recently, an improved single-channel prototype achieved a time resolution of \SI{12.5 +- 0.8}{\pico\second} with a CsI photocathode and a readout pad of \SI{10}{\milli\metre} diameter~\cite{Utrobicic2025SingleChannel}. This development highlighted the importance of detector assembly, electrical connectivity and signal integrity for obtaining the best timing performance.
In addition to induced electrical signals, electron avalanches in suitable gas mixtures produce scintillation light that can be used for optical readout. Fraga et al.\ demonstrated CCD readout of gas electron multiplier (GEM) detectors operated in He/CF$_4$ mixtures, including imaging of the charged particle tracks produced by thermal neutron conversion in $^{3}$He~\cite{Fraga2002CCD}. Measurements of GEM scintillation in He/CF$_4$ and Ar/CF$_4$ mixtures also established the dependence of light yield and spectral distribution on the gas composition~\cite{Fraga2003Scintillation}. The visible component of the emission in CF$_4$-based mixtures makes these gases suitable for readout with imaging sensors. These studies provide a basis for selecting gas mixtures and photon detectors with compatible emission and sensitivity spectra.
Optical readout can provide spatial information without requiring a finely segmented charge collection electrode. Its implementation in conventional Micromegas detectors is, however, restricted by the opaque substrate supporting the anode. This was addressed by integrating a Micromegas amplification structure on a glass substrate with a transparent indium tin oxide (ITO) anode~\cite{glassMicromegas}. Operated in an Ar/CF$_4$ gas mixture, this detector allowed avalanche scintillation light to be recorded with cameras and demonstrated X-ray imaging \cite{Cools2023}. The transparent substrate therefore enables optical access to the amplification region while retaining electronic readout of the anode.
Combining the PICOSEC detection concept with glass Micromegas technology offers the possibility of obtaining complementary timing and position information from electronic and optical signals. Silicon photomultipliers (SiPMs) placed behind the transparent anode can record the temporal response of the scintillation light, while the relative signal amplitudes across an array can be used to reconstruct the particle hit position. The gas mixture must nevertheless satisfy the requirements of both amplification and light detection. Although CF$_4$-based mixtures can provide substantial visible scintillation, this does not directly imply sufficient visible light output from the strongly quenched Ne/CF$_4$/C$_2$H$_6$ mixture used for precise PICOSEC timing. In the present detector, a solid wavelength shifter is therefore integrated on the anode to convert the avalanche-induced ultraviolet scintillation light into visible light detectable by the SiPMs. Alternatively, SiPMs with (V)UV sensitivity may be chosen in combination with a substrate and anode layer with extended UV transparency.

In this work, we investigate SiPM-based optical readout of single-channel PICOSEC glass Micromegas prototypes incorporating a tetraphenyl butadiene (TPB). Measurements with \SI{150}{\giga\electronvolt} muons at the CERN SPS are used to compare the timing response obtained from electronic and optical signals and to evaluate position reconstruction from the signal-amplitudes recorded by a $2 \times 2$ SiPM array. Additionally, camera-based imaging readout is explored as an alternative means of recording the spatial distribution of avalanche light. These measurements assess the potential of combining electronic timing measurements with optical position reconstruction in PICOSEC Micromegas detectors.

\section{Experimental methods}

\begin{figure}[h!]
\centering
\makebox[\linewidth]{\includegraphics[width=0.7\textwidth]{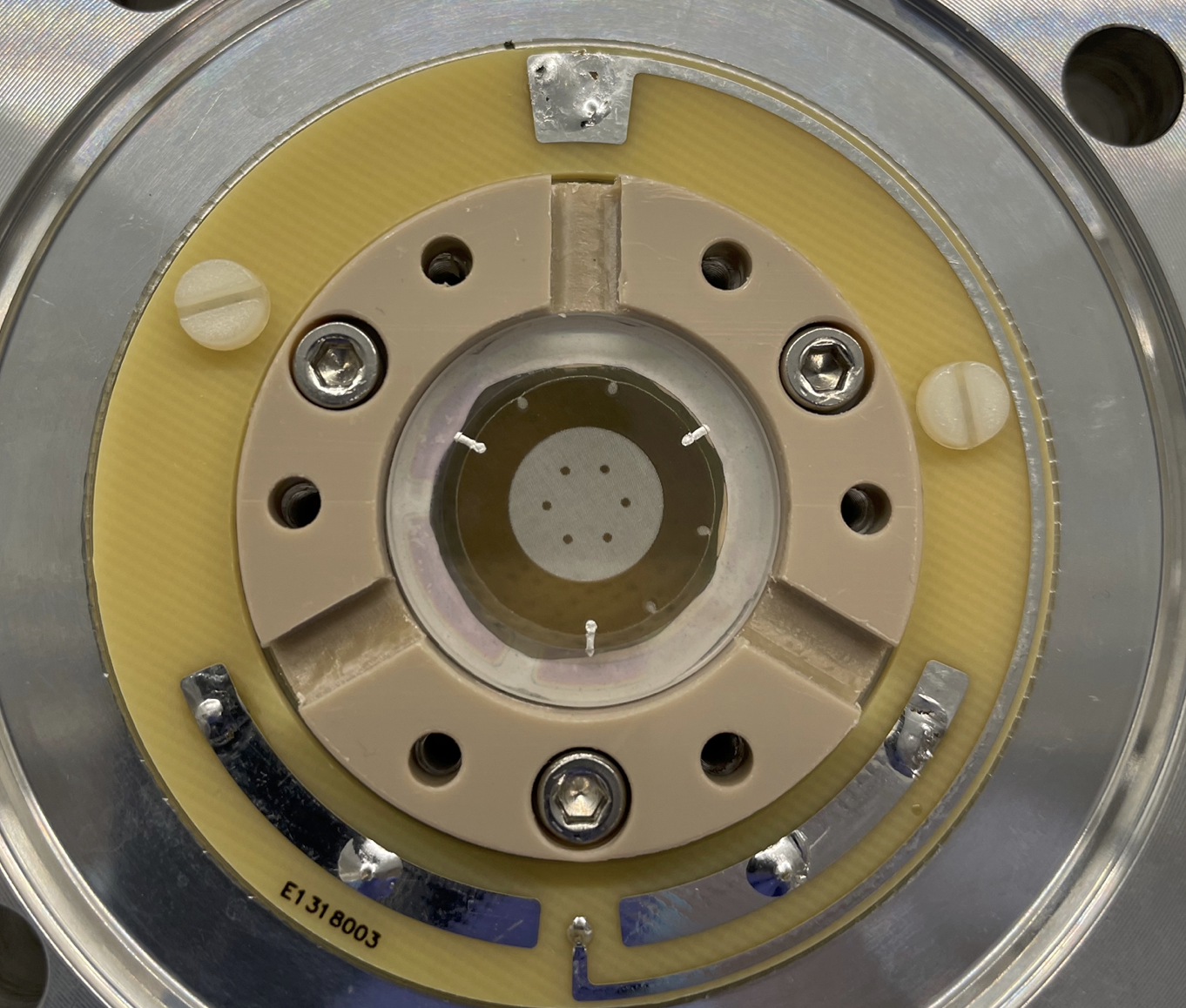}}
\caption{Glass Micromegas glued to vacuum flange. The \SI{10}{\milli\metre} diameter active area with 6 pillars is visible.}
\label{setupImage}
\end{figure}

To enable optical readout of PICOSEC Micromegas, prototypes were integrated on a transparent substrate. Fused silica substrates with a thickness of \SI{3}{\milli\metre} were coated with a patterned layer of indium-tin-oxide (ITO), which acts as electrode while maintaining high optical transparency. Subsequently, a standard bulk Micromegas process was employed to integrate a micro-mesh at a distance of \SI{128}{\micro\metre} above the anode held in place by pillars defined in multiple layers of photoresist. The active area of the detectors is \SI{10}{\milli\metre} and the mesh is supported by 6 pillars. The bulked detectors were glued to the flange of a vacuum chamber as shown in figure \ref{setupImage}. 

\begin{figure}[h!]
\centering
\makebox[\linewidth]{\includegraphics[width=0.7\textwidth]{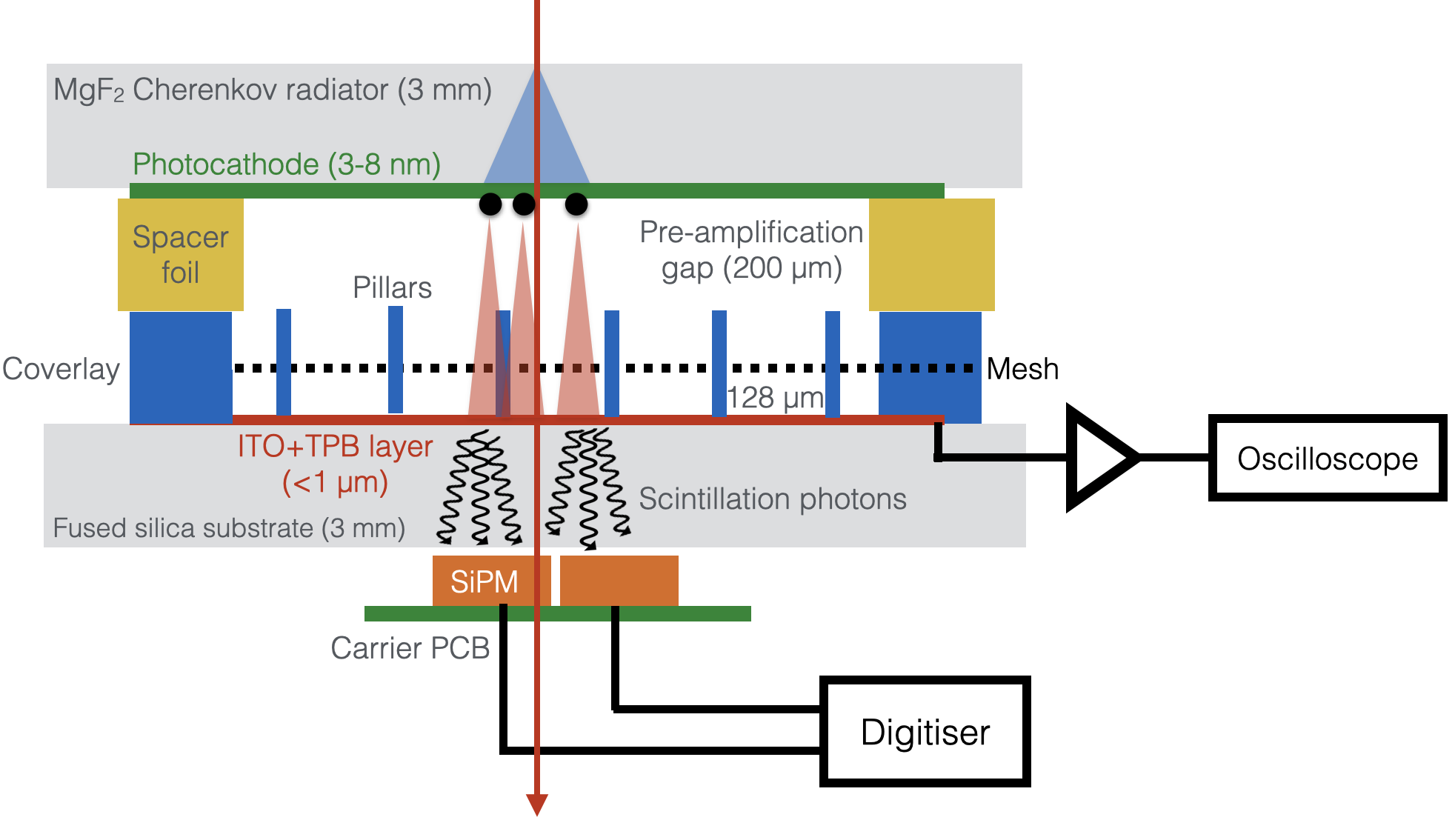}}
\caption{Detector concept cross-section showing Cherenkov radiator, photocathode, indium tin oxide (ITO) + tetraphenyl butadiene (TPB) anode and SiPM placement. The amplification stage consists of a micro-mesh supported by pillars made from Coverlay fixed to the anode. Schematic not drawn to scale.}
\label{setupSchematic}
\end{figure}

A spacer foil made of FR4 with a thickness of \SI{200}{\micro\metre} was placed on top to define the thickness of the drift or pre-amplification region between the micro-mesh and a cathode placed above. A \SI{3}{\milli\metre} thick disk of MgF$_2$ was used as a Cherenkov radiator with a semi-transparent photocathode coated on one side. The spacer foil included Cu traces which allowed to make electrical contact to the anode, mesh and cathode electrodes once the elements were pressed together. A schematic of the cross-section of the detector is shown in figure \ref{setupSchematic}.

The detector was operated with a \SI{3}{\nano\metre} thick Ti layer as metallic photocathode.

The characterisation of the achievable timing and spatial response of the optical PICOSEC Micromegas prototypes was performed during DRD1 test beam campaigns at the CERN SPS H4 beam line with \SI{150}{\giga\electronvolt} muons. The detector was operated in a setup including a GEM-based telescope with three layers of triple-GEMs for accurate position information with a resolution on the order of \SI{100}{\micro\metre}. In addition to the tracking layers, an MCP-PMT was used for triggering and as timing reference. The time resolution of the MCP-PMT (Hamamatsu R3809U-50) was measured to be better than \SI{5}{\pico\second} \cite{Bortfeldt2020a}.
The prototypes were operated in a sealed gas volume built from high vacuum-grade components which allowed for stable operation over multiple days without degradation of the quality of the gas filling. A gas mixture of Ne/CF$_4$/C$_2$H$_6$ (\SI[parse-numbers=false]{80/10/10}{\percent}) was used at a pressure of \SI{1}{\bar}. This gas mixture was chosen due to its good timing resolution achieved in PICOSEC Micromegas measurements. 
Since the gas mixture does not exhibit strong visible light scintillation, a solid wavelength shifter was used to convert UV light from the avalanche to the visible range \cite{IGNARRA20121217} which is recorded by the SiPMs. A layer of tetraphenyl butadiene (TPB) with a thickness of approximately \SI{860}{\nano\metre} was deposited on the anode by thermal evaporation after integration of the detectors. TPB absorbs in the UV range and features a VIS emission peak centered around \SI{420}{nm} \cite{IGNARRA20121217}. Since this material is insulating, no electrical short was created between the micro-mesh and pillars during the deposition process. The evaporation of the TPB through the micromesh may cause an imprint of the mesh on the TPB layer on the anode which may in turn modify the electric field in the amplification region.

Signals from prototypes were read out electronically from the anode as well as optically with an array of SiPMs placed outside of the gas volume facing the anode. An array of 4 SiPMs in a 2x2 arrangement was used to record scintillation light signals. SiPMs with an active area of \SI{3}{\milli\metre} $\times$ \SI{3}{\milli\metre} at a pitch of \SI{5}{\milli\metre} were used with a peak sensitivity at around \SI{425}{\nano\metre} (brand: onsemi, type: MICROFC-30035-SMT-TR, \SI{35}{\micro\metre} microcell size \cite{onsemi_c_series}). 
Unless otherwise noted, timing was extracted offline from recorded waveforms with a \SI{20}{\percent} constant fraction discriminator (CFD) algorithm \cite{Gedcke1967} using a sigmoid fit of the rising edge of the acquired signals. The timestamp of the reference time was extracted in the same way from the MCP-PMT signal.
Background acquisitions with the detector switched off were recorded to confirm that the observed light comes from secondary scintillation in the amplification structure. With the detector switched off, no pulses with comparable signal amplitude were observed.

\section{Results}

The detector was read out electronically with an oscilloscope as well as optically to compare the achievable response. The reported results are from the characterisation of the detector during a muon test beam campaign at the CERN SPS H4 beamline. 

\subsection{Electronic readout of glass Micromegas}

Electronic signals from the anode were acquired with a readout chain made up of a custom-built current sensitive preamplifier \cite{Utrobicic2023Picosec} and digitised by a \SI{1}{\giga\hertz} bandwidth oscilloscope with a sampling rate of \SI{10}{\giga\sample\per\second}. Optical signals from the SiPMs were acquired either with the same oscilloscope or with a \SI{500}{\mega\sample\per\second} digitiser for position resolution studies. No preamplifier or other electronics were used for the SiPM signals as obtained signals were large enough to be recorded directly.

\begin{figure}[h!]
    \centering
    \begin{subfigure}[t]{0.48\textwidth}
        \centering
        \includegraphics[width=\linewidth]{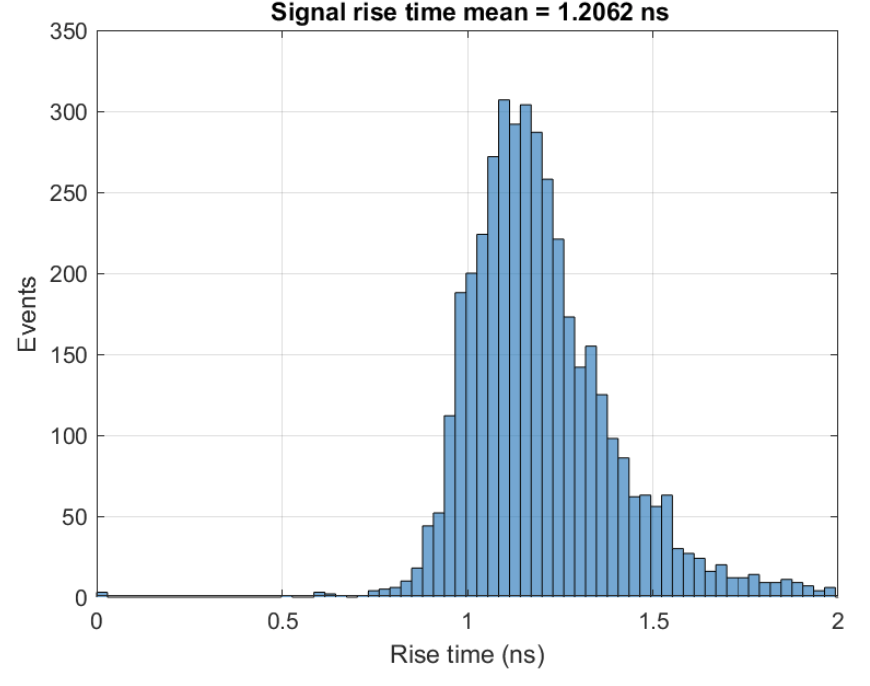}
        \caption{Rise time distribution of signals electronically read out with an oscilloscope. Rise time is measured as time from 10\% to 90\% of signal amplitude.}
        \label{electronicReadoutRiseTime}
    \end{subfigure}
    \hfill
    \begin{subfigure}[t]{0.48\textwidth}
        \centering
        \includegraphics[width=\linewidth]{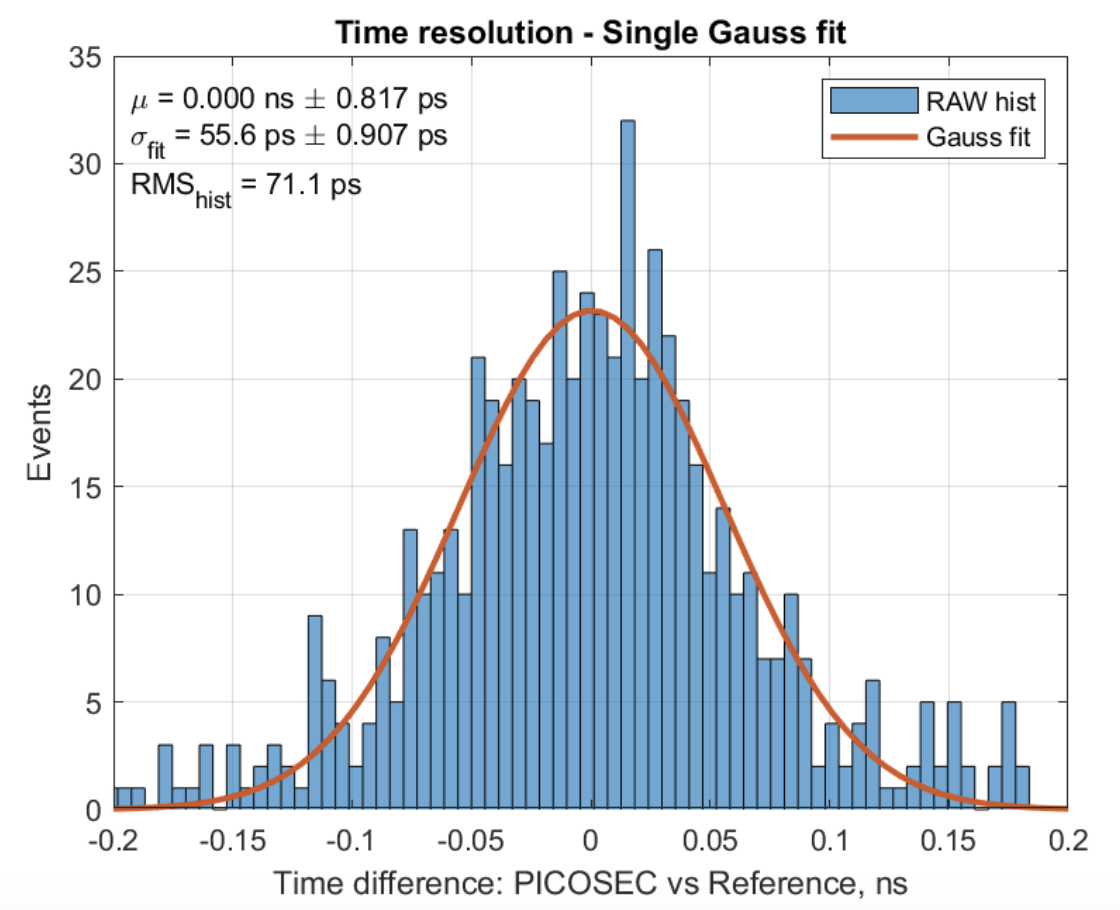}
        \caption{Distribution of signal arrival times determined with
                 a \SI{20}{\percent} CFD algorithm from electronic signals.
                 A time resolution of \SI{55}{\pico\second} is obtained from the
                 $\sigma$ of a Gaussian fit to the distribution.}
        \label{electronicReadoutSAT}
    \end{subfigure}
    \caption{Rise time and signal arrival time distributions for electronic readout. The detector was operated with a Ne/CF$_4$/C$_2$H$_6$ gas mixture. The mesh was grounded and a voltage of \SI{-400}{V} was applied on the cathode. The anode voltage was \SI{400}{V}.}
    \label{fig:electronicReadoutTiming}
\end{figure}

The rise time of electronic signals acquired from the anode was around \SI{1.2}{\nano\second} (\SIrange{10}{90}{\percent}) as shown in figure \ref{electronicReadoutRiseTime}. This compares to approximately \SI{0.7}{\nano\second} from PICOSEC Micromegas prototypes on a PCB substrate with metallic anode. The slower rise time is probably due to the resistivity of the ITO anode layer as well as the connectivity between the anode and the preamplifier which consisted of a metallic pin feedthrough with a length of approximately \SI{2}{\centi\metre}.

The obtained resolution compares to approximately \SI{35}{\pico\second} obtained with comparable PICOSEC Micromegas prototypes with Ti photocathode with conventional electronic readout. This is limited by the low number of photoelectrons from a Ti photocathode. The degraded timing resolution of the glass Micromegas prototypes may be explained by the slower signal rise time and suboptimal signal routing. While the signal is directly routed to a coaxial connector through the PCB in conventional PICOSEC Micromegas prototypes thus minimising the unshielded signal path, in the glass Micromegas prototypes, the unshielded path and simple pin feedthrough may negatively affect signal integrity.

\subsection{Optical SiPM readout}

\begin{figure}[h!]
    \centering
    \begin{subfigure}[t]{0.48\textwidth}
        \centering
        \includegraphics[width=\linewidth]{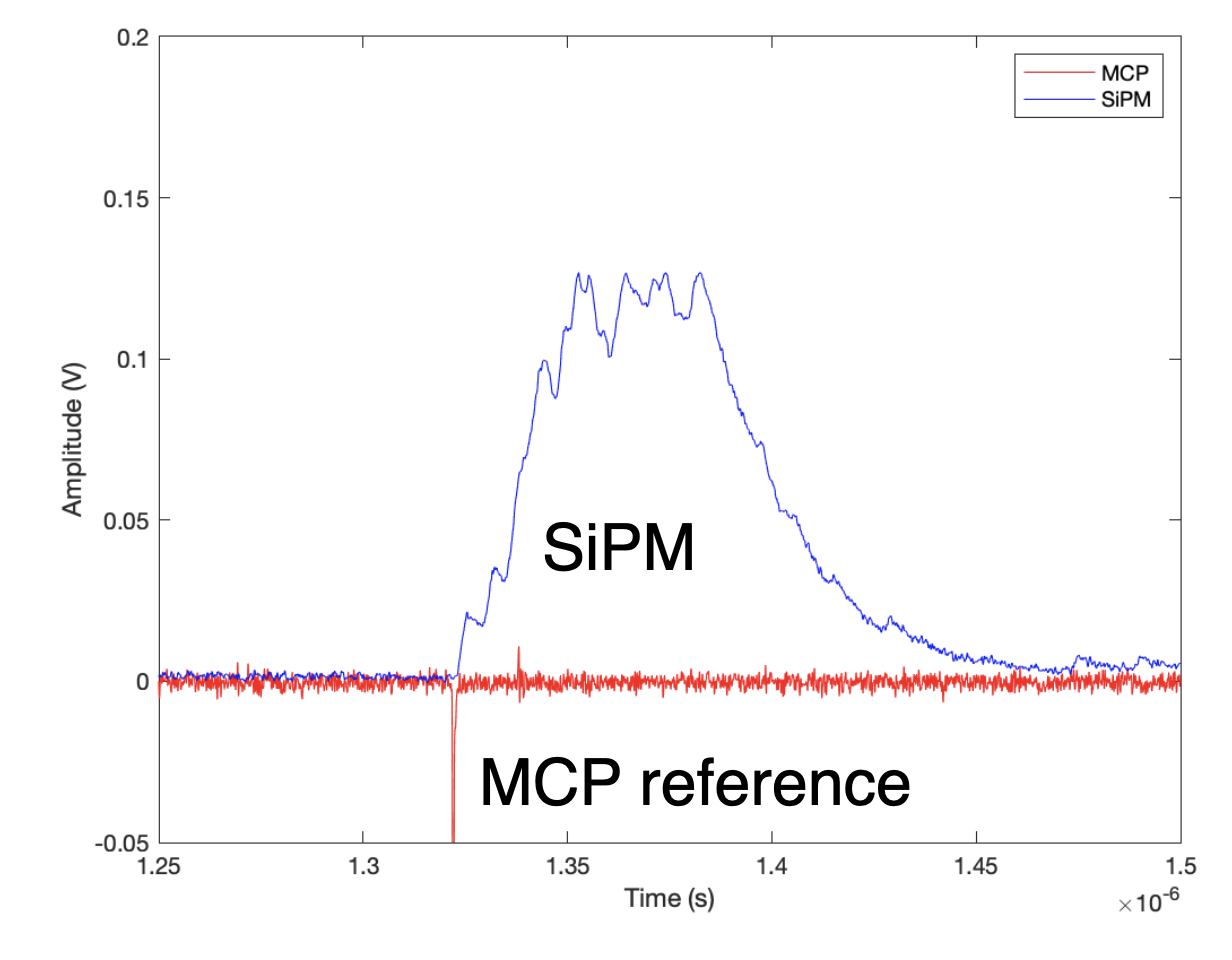}
        \caption{Ar/CF$_4$ (\SI[parse-numbers=false]{80/20}{\percent}) gas mixture. The detector was read out with a Hamamatsu S14160-4050HS SiPM with a bias of \SI{40}{V}. The detector was operated with \SI{560}{V} on the cathode and \SI{520}{V} applied to the anode.}
        \label{sipmWaveformLowGain}
    \end{subfigure}
    \hfill
    \begin{subfigure}[t]{0.48\textwidth}
        \centering
        \includegraphics[width=\linewidth]{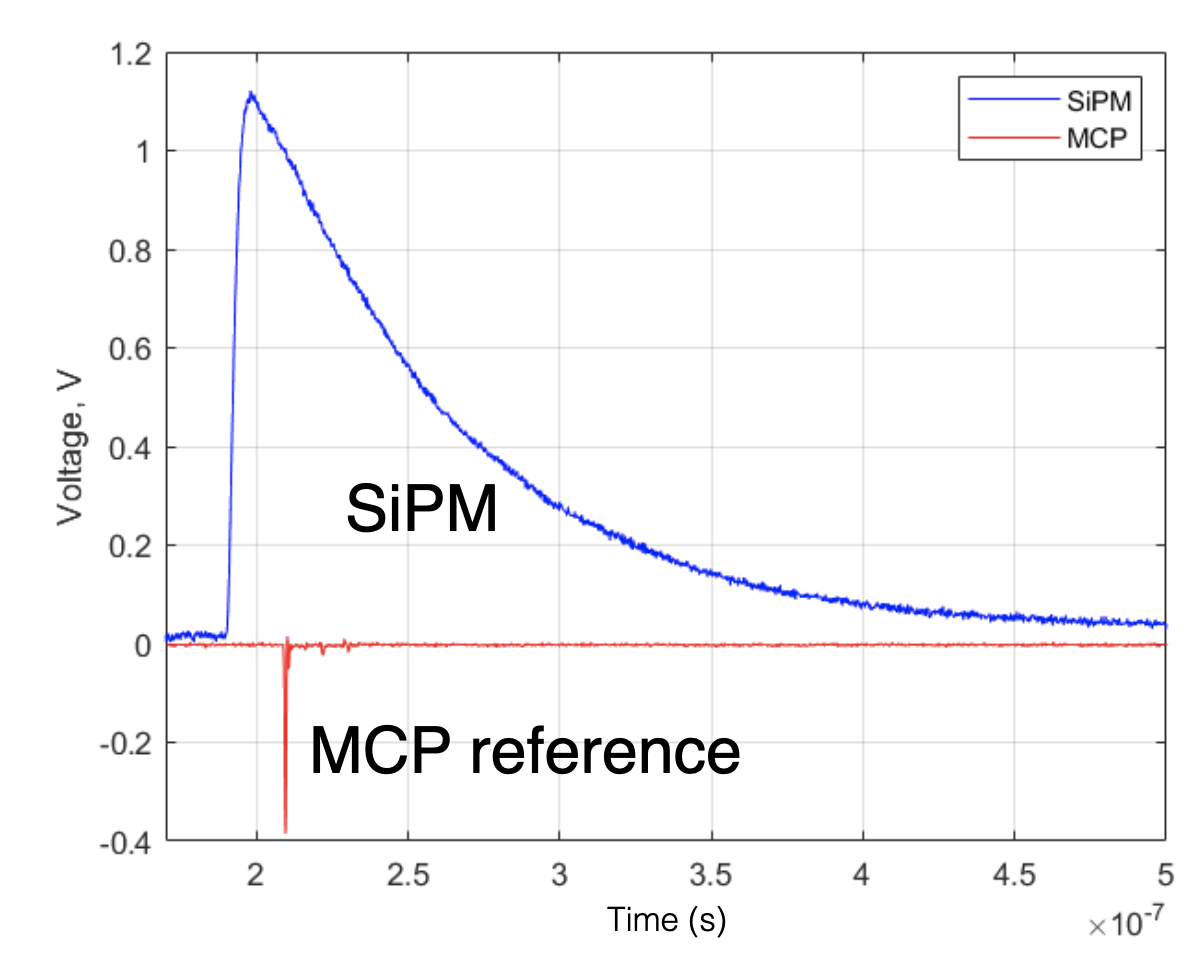}
        \caption{Ne/CF$_4$/C$_2$H$_6$ (\SI[parse-numbers=false]{80/10/10}{\percent}) gas mixture with TPB wavelength shifter. The detector was read out with an onsemi MICROFC-30035-SMT-TR SiPM with a bias of \SI{30}{V}. The detector was operated with \SI{-400}{V} on the cathode and \SI{400}{V} applied to the anode.}
        \label{sipmWaveformHighGain}
    \end{subfigure}
    \caption{Exemplary signals obtained in different gas mixtures. The slow fall time is mainly attributed to the SiPM recovery time. A Ti photocathode was used.}
    \label{fig:sipmWaveforms}
\end{figure}

Optical signals were read out from a single SiPM element of a 2x2 array. SiPMs were placed outside of the gas volume directly in contact with the substrate on which the anode was coated. Depending on the gas mixture and the obtainable gain, different signal waveform characteristics were observed. In an Ar/CF$_4$ (\SI[parse-numbers=false]{80/20}{\percent}) gas mixture, the achievable gain was relatively low and the resulting signals showed the arrival of individual photons as shown in figure \ref{sipmWaveformLowGain}. In comparison, a Ne/CF$_4$/C$_2$H$_6$ (\SI[parse-numbers=false]{80/10/10}{\percent}) gas mixture achieved higher gain. The additional use of a TPB layer as solid wavelength shifter resulted in signal waveforms with significantly higher amplitude as shown in figure \ref{sipmWaveformHighGain}. These higher gain waveforms are smoother due to the larger number of photons recorded.

\begin{figure}[h!]
    \centering
    \begin{subfigure}[t]{0.48\textwidth}
        \centering
        \includegraphics[width=\linewidth]{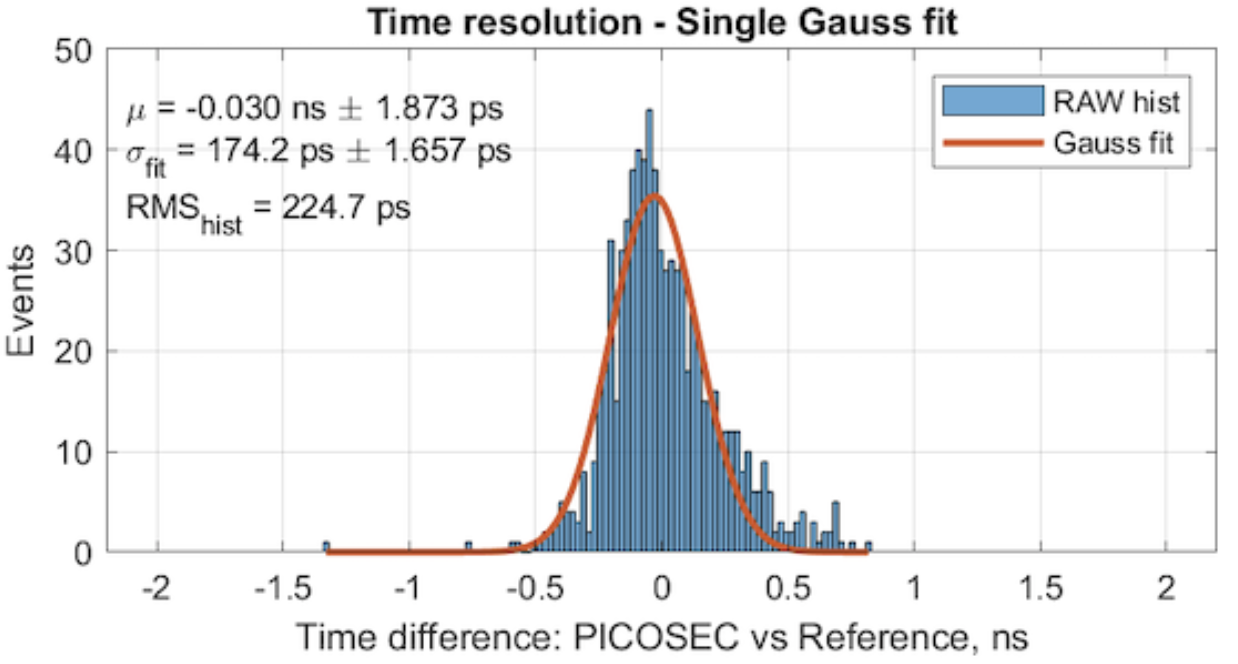}
        \caption{Time resolution extracted with a \SI{20}{\percent} CFD algorithm.}
        \label{opticalReadoutResolutionCFD}
    \end{subfigure}
    \hfill
    \begin{subfigure}[t]{0.48\textwidth}
        \centering
        \includegraphics[width=\linewidth]{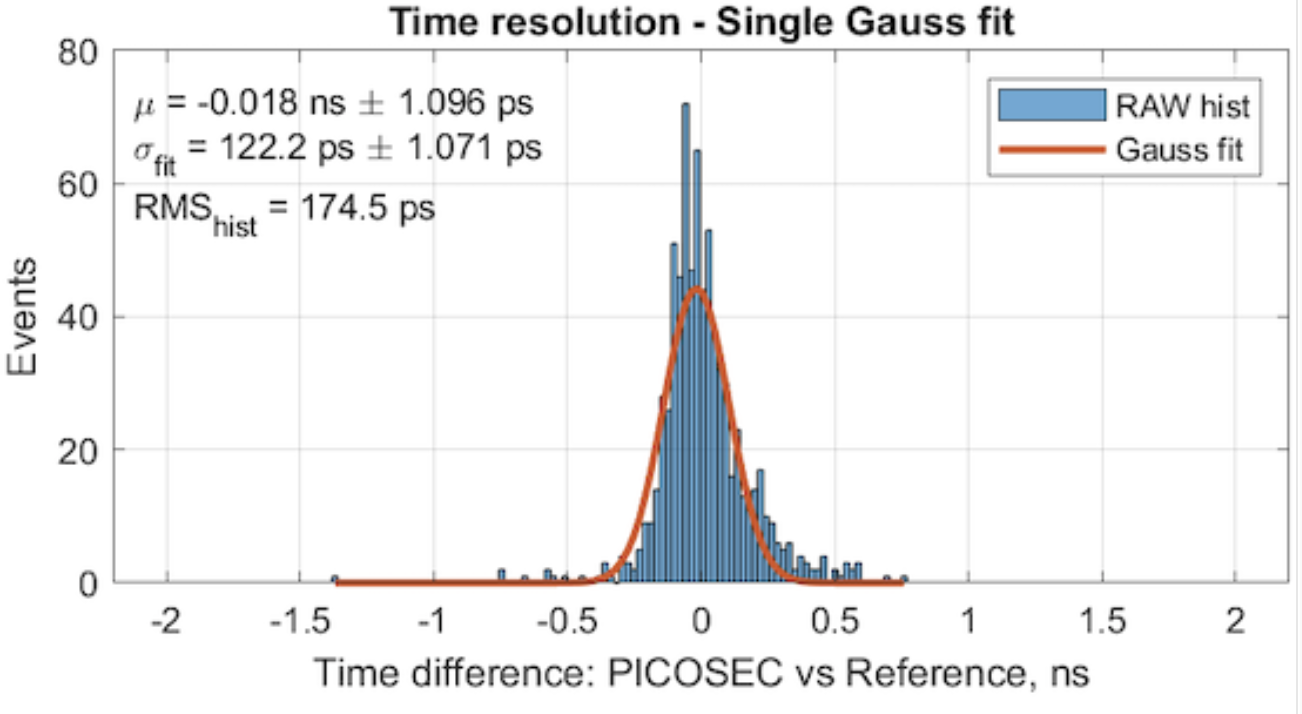}
        \caption{Time resolution extracted with a constant threshold
                 of \SI{100}{\milli\volt}.}
        \label{opticalReadoutResolutionTH}
    \end{subfigure}
    \caption{Time resolution obtained with different timing algorithms. A Ne/CF$_4$/C$_2$H$_6$ (\SI[parse-numbers=false]{80/10/10}{\percent}) gas mixture was used with \SI{-400}{V} on the cathode and \SI{400}{V} applied to the anode.}
    \label{fig:opticalReadoutResolution}
\end{figure}

The time resolution was extracted by determining a timestamp of the signal waveform and calculating the time difference between the signal timestamp and the reference MCP-PMT signal. A \SI{20}{\percent} CFD algorithm was compared with a fixed threshold based timing algorithm with a threshold set at \SI{100}{\milli\volt}. In both cases, a correction for the residual time walk was applied. A comparison of the timing resolution of the two methods is shown in figure \ref{fig:opticalReadoutResolution}.

\begin{figure}[h!]
\centering
\makebox[\linewidth]{\includegraphics[width=0.7\textwidth]{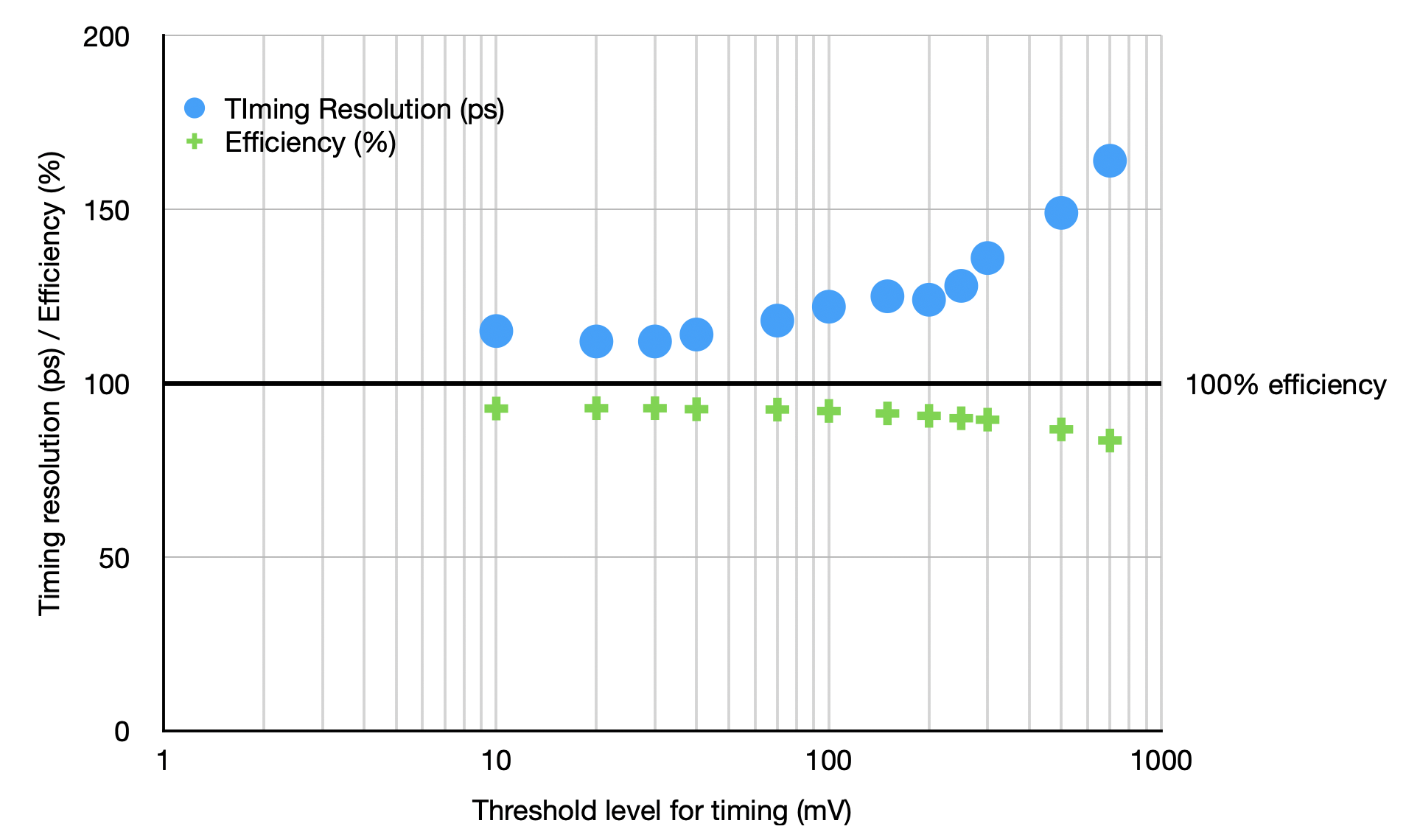}}
\caption{Dependence of time resolution and efficiency on threshold value. Same configuration and voltages as listed in figure \ref{fig:opticalReadoutResolution}.}
\label{efficiencyResolutionThresholdScan}
\end{figure}

The choice of the threshold value impacts timing resolution as well as efficiency. A comparison of the achievable time resolution and efficiency as a function of the threshold value is shown in figure \ref{efficiencyResolutionThresholdScan}.

For low threshold values from \SIrange{10}{100}{\milli\volt}, good timing resolution around \SI{120}{\pico\second} is achieved while the efficiency remains high above \SI{90}{\percent}.

The time resolution achieved with optical readout is lower than the one achieved with electronic readout which can be a result of the slow scintillation mechanism of CF$_4$ as well as the readout of the SiPMs. Subtracting the electronic time resolution from the one obtained with SiPM readout, an additional jitter of approximately \SI{110}{ps} from the optical readout chain is obtained.

\begin{figure}[h!]
\centering
\makebox[\linewidth]{\includegraphics[width=0.7\textwidth]{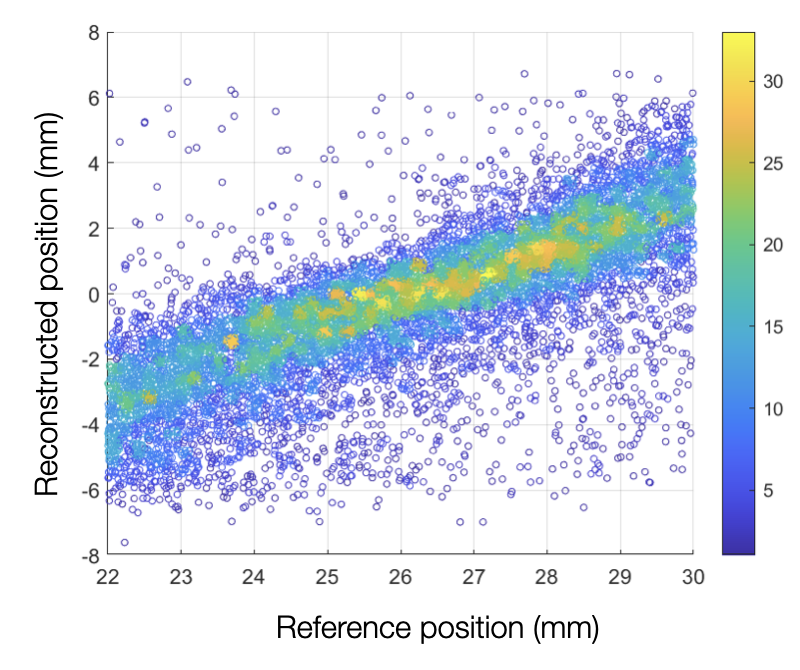}}
\caption{Correlation of reconstructed hit location and reference position from the tracker for X coordinate.Same configuration and voltages as listed in figure \ref{fig:opticalReadoutResolution}.}
\label{positionReconstruction}
\end{figure}

In addition to extracting time resolution from SiPM waveforms, the amplitude distribution on multiple SiPM channels can be used to reconstruct the particle hit location. The signals from a 2x2 array of SiPMs were read out with a digitiser and a center-of-gravity algorithm was used to offline reconstruct hit locations. Reconstructed hit locations were then compared to the reference position obtained from the tracking telescope. A correlation of the reconstructed position from the SiPM amplitudes and the reference position from the tracker is shown in figure \ref{positionReconstruction}. At the lower and upper limits of the shown range, the distribution differs from a linear relationship and widens which can be explained by the limited area covered by the 2x2 array of SiPMs.

\begin{figure}[h!]
\centering
\makebox[\linewidth]{\includegraphics[width=0.7\textwidth]{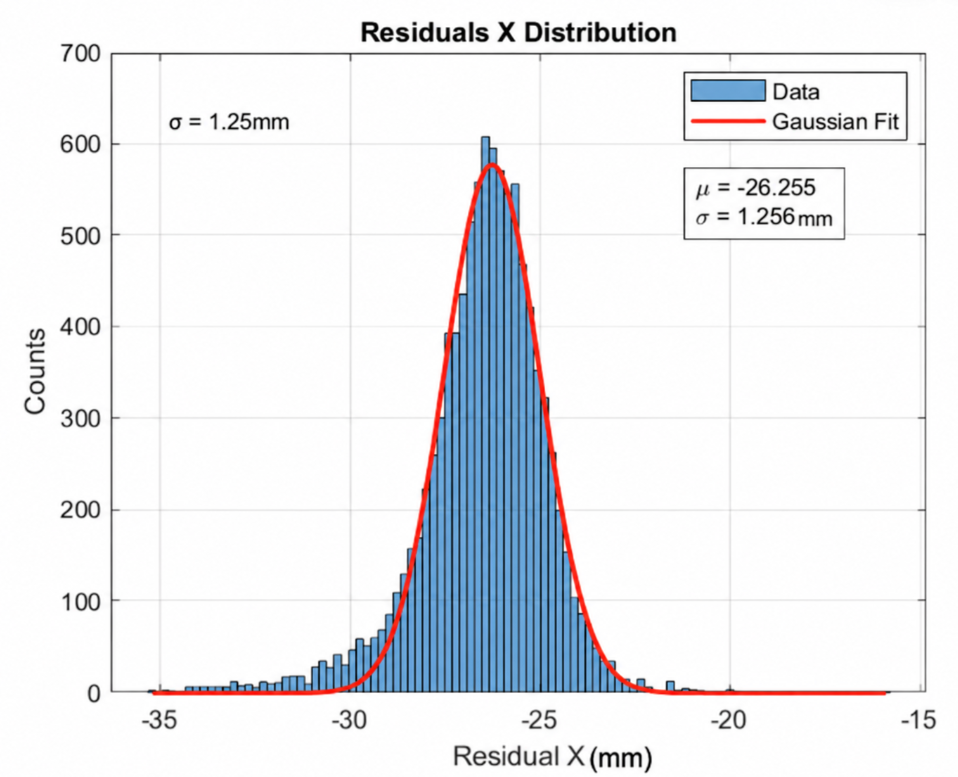}}
\caption{Residuals between reconstructed hit location and reference position as measured by the tracker telescope \cite{Bortfeldt2018Picosec} in X coordinate. A spatial resolution of $\sigma = \SI{1.25}{\milli\metre}$ was achieved. Same configuration and voltages as listed in figure \ref{fig:opticalReadoutResolution}. The tail on the left may be due to non-linearities as shown on the edges of the correlation curve in figure \ref{positionReconstruction}.}
\label{positionReconstructionResiduals}
\end{figure}

The difference between the reconstructed and tracker position was then used as residual to create a histogram of residuals in X and Y directions as shown in figure \ref{positionReconstructionResiduals}. The width of a Gaussian fit to the residuals distribution was then used as a measure of the spatial resolution and a value of $\sigma = \SI{1.25}{\milli\metre}$ in X direction was obtained. 

The obtained time resolution is slightly better than the resolution expected for binary readout. Simulations show that a significant limiting factor is the distance of the SiPMs from the location of scintillation light emission as well as their size.

\subsection{Camera readout of PICOSEC glass Micromegas}

\begin{figure}[h!]
\centering
\makebox[\linewidth]{\includegraphics[width=0.7\textwidth]{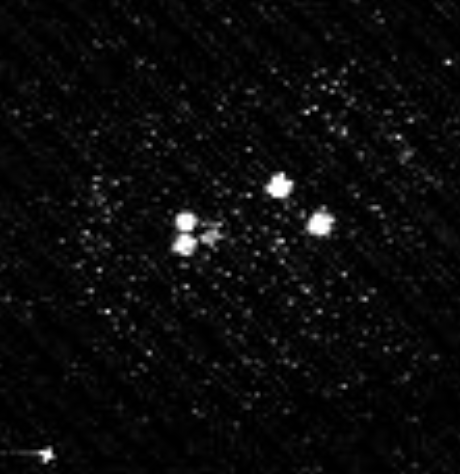}}
\caption{Image of glass Micromegas response to incident muons obtained with EMCCD camera. A Ne/CF$_4$/C$_2$H$_6$ (\SI[parse-numbers=false]{80/10/10}{\percent}) gas mixture was used with \SI{-400}{V} on the cathode and \SI{450}{V} applied to the anode.}
\label{sipmCameraFrame}
\end{figure}

As an alternative to optical readout with SiPMs, the glass Micromegas were read out with an electron multiplying (EMCCD) camera. A Hamamatsu ImageEM camera with a resolution of 512 x 512 pixels \cite{EMCCD} was used to acquire images of the response of the detector to muons from the same test beam as described above. An exemplary image obtained by the camera readout is shown in figure \ref{sipmCameraFrame}. 

The obtained image was obtained with the maximum EM gain of 1200x with 2 x 2 binning and \SI{7}{\milli\second} exposure time. The detector was operated with a voltage of \SI{-400}{\volt} on the cathode and \SI[retain-explicit-plus = true]{+450}{\volt} on the anode. A Ti photocathode was used.
The signal spots display distributed avalanches across the active area of the detector.No synchronisation of the camera readout with the tracker as hit position reference was achieved. Using a photocathode with a higher quantum efficiency such as CsI would result in a larger number of signal spots per particle and thus allow better determination of the original hit location.

\section{Conclusions}

Combined optical and electronic readout of PICOSEC Micromegas precise timing detectors can be used to obtain tens of \si{\pico\second} timing resolution and millimetre-scale position resolution. Coupling a SiPM array to a PICOSEC glass Micromegas enables readout of scintillation light produced during avalanche multiplication. High light yield is obtained in a Ne/CF$_4$/C$_2$H$_6$ gas mixture with the use of tetraphenyl butadiene (TPB) as wavelength shifter on the anode. The TPB was not patterned and no signs of charging up behavior were observed. While the combination of this gas mixture with the TPB wavelength shifter was shown to work, TPB is known to degrade under extended irradiation and ion bombardment. Therefore, alternative wavelength shifter materials should be examined to extend detector lifetime.
The time resolution extracted from optical signals is worse than the one obtained from electronic signals. This may be caused by the slow scintillation time characteristics of CF$_4$ or the decay time of the wavelength shifter which are both on the order of nanoseconds to tens of nanoseconds. A faster scintillating mixture or wavelength shifter may result in a sharper optical signal rise time and consequently improved timing resolution. Fixed threshold timing with time walk correction was shown to perform better than a CFD algorithm on the SiPM waveforms with lower threshold values giving the best resolution values.
Finally, optical readout with a camera imaging scintillation light is demonstrated which may enable fine position reconstruction for events with a large number of photoelectrons.


\bibliographystyle{JHEP}
 \bibliography{biblio.bib}

\end{document}